\documentclass[onecolumn,amsmath,amssymb]{revtex4}

\usepackage[dvipdfmx]{graphicx}
\usepackage{dcolumn}
\usepackage{bm,color}
\usepackage{amsmath}

\begin{document}

\preprint{APS/}

\title[Solitons ...]{Solitons in nuclear time-dependent density functional theory} 

\author{$^1$Yoritaka Iwata}
 \email{iwata_phys@08.alumni.u-tokyo.ac.jp}

\affiliation{$^{1}$ Faculty of Chemistry, Materials and Bioengineering, Kansai University, Osaka 564-8680, Japan. }

\maketitle

The soliton existence in sub-atomic many-nucleon systems is discussed.
In many nucleon dynamics represented by the nuclear time-dependent density functional formalism, much attention is paid to energy and mass dependence of the soliton existence.
In conclusion, the existence of nuclear soliton is clarified if the temperature of nuclear system is from 10 to 30 MeV.
With respect to the mass dependence $^{4}$He and $^{16}$O are suggested to be the candidates for the self-bound states exhibiting the property of nuclear soliton. \\

\section{Introduction}
The concept of nuclear soliton is proposed by its existence in the three-dimensional nuclear time-dependent density functional formalism.
The solitons in this article are the waves stably traveling without changing the shape and velocity even after collisions between waves (Fig.~1).
In this sense, as for the terminologies of classical and quantum field theory, what we study in this article is not similar to the topological soliton [Ma04, We12], but rather corresponding to the non-topological soliton [Le92].
In the following we refer simply to ``soliton'' for a kind of non-topological soliton.
The mathematically common property of soliton (for example, see [Ab11]) has been clarified as
\begin{itemize}
\item nonlinearity
\item dispersive property
\end{itemize}
being independent of the size and medium of wave. 
The common properties of solitons are essential to the soliton existence, and several uncommon properties specific to nuclear soliton such as
\begin{itemize}
\item quantum effect with the fermi statistics
\item many-body effect leading to the collectivity
\end{itemize}
can modify the conditions of soliton existence, where a competition between them possibly appears.
In the most of preceding soliton researches, the size and model dependent additional properties are not seriously taken into account.
Here we employ the nuclear time-dependent density functional theory (TDDFT) in which all the above four properties are included in a self-consistent manner.
In particular the collectivity of many-nucleon systems has been successfully treated by the nuclear DFT with and without time-dependence (for example, see [Gr96]).

The solitons are observed in any scales, if the mathematically common property is held by the master equation.
This fact is something to do with the size and model dependence of the two common property.
The nuclear soliton is found in sub-atomic  femto-meter scales whose energy is at the order of MeV (mega electron volt). 
Such specific scale arises from the effective unit of motion: the nucleon degree of freedom in case of nuclear soliton.
For example, the effective unit of motion for the optical soliton is the photon.
In other words, as is known in the nuclear physics, the motion of nucleus at the energy order of MeV is governed by the independent nucleon motion (for example, see [Ri80]).

The soliton is a wave with both individuality and stability. 
On the other hand, the nuclear soliton is also regarded as (bringing about) a state of nuclear matter: the perfect fluidic state.
It is worth mentioning here that the perfect fluidity can be rephrased as the inertness in the context of reaction theory.
Accordingly the nuclear soliton is expected to be associated with some important physics if its existence is established.
Indeed, the perfect fluidity leads to the conservation of the number of vortex. 
Since celestial bodies consist of nuclear matter, the quantitative understandings of the nuclear soliton is possible to show a new aspect for the matter/heat transportation inside the (compact) stars. 
Furthermore the perfect fluidity is associated with the dissipation property of low-energy heavy-ion collisions that has been a long standing open problem in the microscopic nuclear reaction theory.
The perfect fluidity is also associated with the conservation of nuclear matter without the loss of any information: i.e., isentropic property arising from the time-reversal symmetry [Iw19].
As the conservation property of soliton has already been utilized in the optical fiber, the preservation property of  nuclear matter is expected to be utilized to the nuclear engineering for preserving and condensing a certain projectile nucleus.
In particular the well preserved nuclear matter is expected to be used for the reduction of the nuclear wastes by the nuclear transmutation with the extremely high intensity/density projectile of reactions, which is not only for making high intensity/density beam but also high projectile-density matter in the nuclear reactor. 

This article is organized as follows.
The basic concepts of wave propagation is introduced in Sec.~\ref{sec2}.
The general definition of solitary wave and soliton is shown in Sec.~\ref{sec3}.
The existence of nuclear soliton is discussed in Sec.~\ref{sec4}.
The summary and perspectives are presented in Sec.~\ref{sec5}.

\begin{figure}[tb]
\begin{center} \label{fig:2}
\includegraphics[width=10cm]{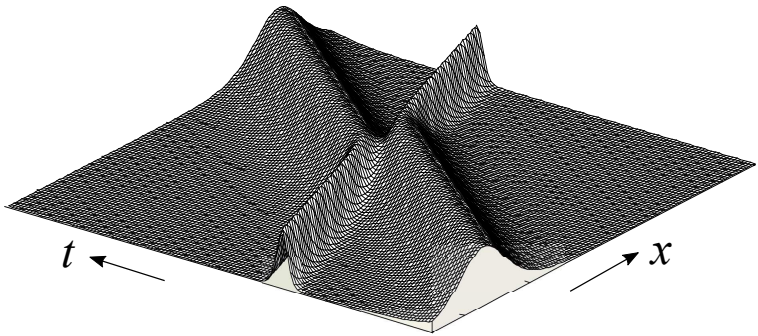} \vspace{5mm}  
\end{center}
\caption{(Color online) Two soliton solution of KdV equation ($\alpha = 1$).
Although the momentum and shape is exactly conserved, time delay appears due to the collision (around $(t,x) = (0,0)$).}
\end{figure}

\section{Equation of waves} \label{sec2}
This section is devoted to introduce the basic concepts for wave propagation, which provides a working area of the soliton research. 
For the purpose of introducing the concept of dispersive property, we begin with  the linear wave:
\begin{equation} \label{eq01}
u(t,x) = A \exp( i (kx - \omega t +\alpha)), 
\end{equation}
in one-dimensional space, where $k$ means the wave number, $\omega$ the angular frequency and $\alpha$ the phase.
This wave is also referred to the plane wave in the multi-dimensional case, and to a traveling wave in more general fields.
The first order linear hyperbolic equation (advection equation) is written by
\begin{equation} 
\partial_t u + c \partial_x u = 0
\label{eq02}
\end{equation}
in one-dimensional space ${\mathbb R}$, where $c$ is a real constant meaning the propagation speed.
It is well known that this equation holds the solution represented by the d'Alembert's formula, so that the plane wave (\ref{eq01}) satisfies this equation.
The linear dispersion relation $\omega = c k$ is satisfied by the plane wave solution. 
The plane wave solution can also be associated with the second order linear wave equations, with respect more closely to the present interest, the Klein-Gordon equation:
\begin{equation} 
\partial_t^2 u - c^2 \partial_x^2 u  +  \left(\frac{mc^2}{\hbar} \right)^2 u = 0
\label{eq03}
\end{equation}
describing a quantum scalar or pseudoscalar fields.
By considering the same plane wave solution, another relation 
$\omega^2  = c^2 (k^2 + m^2 c^2/\hbar^2) $
is obtained, which is asymptotically equal to $\omega  = \pm c k$  (Fig.~2).
Note that the dispersion relation in the massless case ($m=0$) also becomes $\omega  = \pm c k$. 

The Schr\"{o}dinger equation is known to describe the non-relativistic quantum physics.
The linear dispersion relation $\omega  = c k$ is violated in case of Schr\"{o}dinger type waves.
On the other hand, it is readily confirmed that the plane-wave solution is also the solution of linear  Schr\"{o}dinger equation:
\begin{equation}
i \partial_t u +  c \partial_x^2 u  =0
\label{eq04}
\end{equation}
in one-dimensional space ${\mathbb R}$, where $c$ is a real constant being represented by $c = -\hbar/2m$ using the Dirac constant $\hbar$ and the mass $m$. 
In this case another dispersion relation $\omega = c k^2$ is satisfied instead.
Such waves without satisfying the linear dispersion relation $\omega = c k$ are called the dispersive wave.
It is worth noting here that the non-relativistic approximation of  Klein-Gordon equation corresponds to the Schr\"{o}dinger equation.
As a result the Schr\"{o}dinger equation is a typical example of dispersive wave equations.

\begin{figure}[tb]
\begin{center}
\includegraphics[width=10cm]{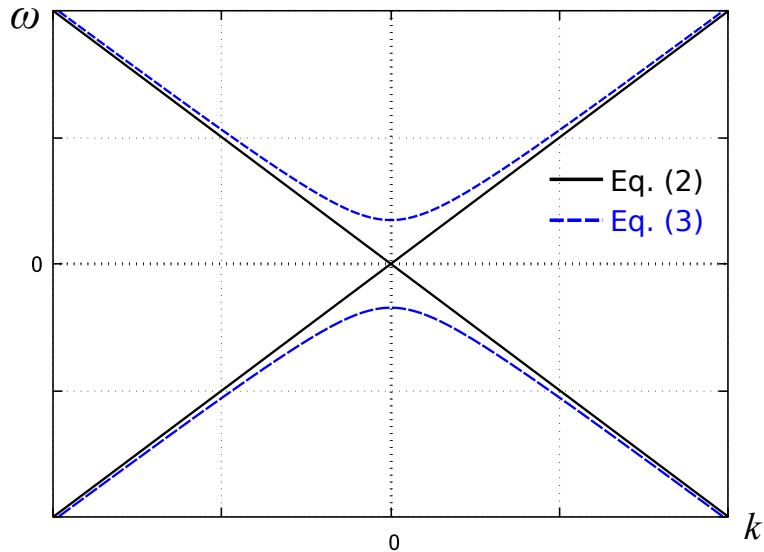}
\end{center}
\caption{(Color online) Dispersion relation associated with Eqs. (\ref{eq02}) and (\ref{eq03}).}
 \label{fig:1}
\end{figure}

\section{Nonlinear dispersive waves}  \label{sec3}

\subsection{Korteweg-de Vries equation}
The concepts of solitary wave and soliton are introduced.
For verifying the soliton existence in sub-atomic quantum equations, we focus on two relevant equations: Korteweg-de Vries equation and nonlinear Schr\"{o}dinger equation.
These equations are not only the dispersive wave equations but also nonlinear evolution equations.

First of all, in the flow of shallow water, the concept of solitary wave was introduced by Scott-Russel [Sc48] in 1844.
Indeed they observe
\begin{itemize}
\item a single wave moves stably on the flat surface without changing the shape and velocity.
\end{itemize}
This is the essential property of solitary wave.
Here the single wave means the wave without undergoing any collisions with the other waves.
Although such a property is common in linear cases, this should not be true in nonlinear cases. 
If solitary waves preserve their shapes and speeds after a collision, the solitary waves holding a transparency is called the soliton. 
In particular, the terminology ”soliton” is introduced by Zabusky and Kruskal in 1965 [Za65].
Indeed, for the initially given sine waves, they are split into several solitary waves, and 
\begin{itemize}
\item the solitary wave moves stably by preserving momentum and shape even after the collisions; 
\item the solitary wave possibly experiences the phase shift and the time delay during the collision;
\end{itemize}
these are the properties to be satisfied by the soliton wave.
That is, the solitary wave is called soliton if it satisfies the above properties.
The transparency leading to the individuality is often called the particle-like property in the soliton theory.
In particular, by comparing the soliton waves before and after the collision, there is not any changes for the momentum and shape, but for the phase.

The equations holding the soliton as a solution are called the soliton equation, and the Korteweg-de-Vries equation (KdV equation, for short) is known as a soliton equation.
In a mathematical sense the concept of solitary wave has been initially studied by the KdV equation [Ko95]
\begin{equation}
 \partial_t u + \alpha u \partial_x u +  \partial_x^3 u  =0,   \quad x \in {\mathbb R},
\label{eq05}
\end{equation}
where $\alpha$ is a real constant.
In the second term $\alpha u$ plays a role of propagation speed (cf. Eq.~(\ref{eq02})), so that the propagation speed depends on the state of wave.
This nonlinear equation is modeling the shallow water waves including both the nonlinearity and the dispersive property, but the dissipation leading to the non-unitary time evolution.
It is worth noting here that the KdV equation is obtained by approximating the imcompressible Navier-Stokes equation (for example, see [La80]).

The plane wave (\ref{eq01}) can be the solution at small amplitude oscillation limit, and then $\omega = c k - k^3$ is approximately satisfied.  
On the other hand KdV equation admits some exact traveling wave solutions:
\begin{equation}
u = \frac{3c}{\alpha} {\rm sech} ^2 \left[ \frac{\sqrt{c}}{2} (x - c t) \right]
\label{eq06}
\end{equation}
where $c$ means the speed of wave propagation.
It is remarkable that Eq.~(\ref{eq06}) holds the form of d'Alembert's solution for the wave equation.
This solution corresponds to the solitary wave solution (one-soliton solution) whose amplitude depends on the propagation speed $c$. 
The solitary wave solution can hold the soliton property that has been examined by obtaining the exact two-soliton solution (Fig.~1).
\begin{equation}
u = \frac{72}{\alpha} ~ \frac{3+4 {\rm cosh}(2x-8t) + {\rm cosh}(4x-64t) }{\{ 3 {\rm cosh}(x-28t) +  {\rm cosh}(3x-36t) \}^2}
\label{eq07}
\end{equation}
asymptotically equal to the superposition of two solitons for large $t$ 
\begin{equation}
u = \frac{12 \kappa_i}{\alpha} {\rm sech} ^2 \left[ \kappa_i (x - 4 \kappa_i^2 t) + \delta_i \right],
\label{eq08}
\end{equation}
where $i=1,2$, $\kappa_1 = 1$, $\kappa_2 = 2$, and $\delta_i$ are constants.
The existence of two-soliton solution ensures the existence of soliton in a given theoretical framework.
In several equations the two-soliton solutions are extended to $N$-soliton solutions (for example, see [Sc73]).

\subsection{Nonlinear Schr\"{o}dinger equation}
A typical soliton equation for non-relativistic quantum dynamics is the nonlinear Schr\"{o}dinger equation (NLS equation, for short).
It reads
\begin{equation}
i \partial_t u +  \partial_x^2 u + k |u|^2 u =0,
\label{eq09}
\end{equation}
where a real number $k$ means the interaction constant, and also the height/depth of potential hill/well.
Indeed, in case of positive $k$, $V(u) = - k |u|^2$ provides a potential well.
Indeed, it holds a solution 
\begin{align*}
u(t,x) = \sqrt{\frac{u_e^2 - 2 u_e u_c} {2k} } ~ {\rm sech} \left[  \sqrt{\frac{u_e^2 - 2 u_e u_c} {4} } (x- u_c t) \right] \exp \left[ i (u_0/2) (x- u_c t) \right] ,
\end{align*} 
where the amplitude of $u$ depends on the constant $k$, which is a specific feature arising from the angular speed $u_c$ and the wave propagation speed $u_e$.
Contrary to the previous KdV equation, the amplitude is proportional to $k^{-1/2}$ and $u_e$. 
Consequently the two factors have been considered to be essential to the soliton propagation: the dispersive property and the nonlinearity.

\subsection{Sturm-Liouville formalism}
Following P. D. Lax [La68], the relation between the KdV and the Schr\"{o}dinger type equations are understood by a simplified Sturm Liouville equation:
\begin{equation}
Ly :=  \partial_x^2 y - U(x,t) y = \lambda y,
\end{equation}
where the periodic boundary condition is imposed, for instance.
This equation can be regarded as the Schr\"{o}dinger equation with the potential $ -\lambda + U(x,t)$.
It is readily seen that
\begin{equation}
\partial_t  ( L y ) = (\partial_t   L) y +  L (\partial_t  y ) 
= (\partial_t   \lambda) y +  \lambda (\partial_t  y )
= - (\partial_t U(x,t)) y + ( \partial_x^2   - U(x,t) ) (\partial_t y)
\label{eq10}
\end{equation}
leads to
\begin{equation}
 (\partial_t L) y  =   - (\partial_t U(x,t)) y.
\label{eq11}
\end{equation}
If $t$-independence of parameter $\lambda$: $\partial_t \lambda = 0$ is further assumed, 
\begin{equation} \begin{array}{ll}
(\partial_t   \lambda) y 
= - (\partial_t U(x,t)) y +L (\partial_t y)  -   (\partial_t  \lambda y ) \\
\quad = - (\partial_t U(x,t)) y +L (\partial_t y)  -   (\partial_t  L y )  \\
\quad = - (\partial_t U(x,t)) y + [ L, \partial_t] y 
\label{eq12}
\end{array} \end{equation}
is obtained, where $[\cdot, \cdot]$ denotes the commutator product.
After generalizing this equation as
\begin{equation} \begin{array}{ll}
(\partial_t   \lambda) y  = - (\partial_t U(x,t)) y + [ L, D] y, 
\label{eq13}
\end{array} \end{equation}
the KdV equation with the potential $U$ and $\alpha = -6$ is obtained by $ - (\partial_t U(x,t))  + [ L, D] = 0$ with $D = f \partial_x^3 + g  \partial _x + h$, $g = -3U f/2$ and $h=-3(\partial_x u) f/4$.
Consequently KdV and Schr\"{o}dinger equations are associated not only by having a soliton solution, but by holding a common mathematical structure.
It is interesting to remind here that the relation between Schr\"{o}dinger, Heisenberg and interaction pictures in quantum field theory (for a textbook, see [Fe03]).

\section{Solitons in nuclear TDDFT}  \label{sec4}
\subsection{Many-nucleon system}
Atomic nucleus is a finite-body many-nucleon system consisting of nucleons: protons and neutrons.
Proton number ranges from 1 to 120 (at the present), and neutron number from 1 roughly to 200.
There expected to exist almost 300 stable nuclei in nature, and the theoretical calculations such as nuclear density functional calculations simulate those nuclei being sufficiently comparable to the experiments.

We are interested in the soliton propagation at the scale of atomic nuclei.
The size of one nucleus ranges from  $10^{-15}$m to $10^{-13}$m, and the corresponding energy is below several 10s of MeV per nucleon.
One of the unique feature of many-nucleon system is found in their finite-body property, which is quite different from most of many-electron systems being treated as infinite matter.
This feature brings about the fact that the self-bound state (the localized wave) is naturally realized in both nature and theory of many-nucleon systems.
Following the general usage of low-energy nuclear physics, the terminology ``low-energy'' is used for the energy below 30 MeV per nucleon.
The relativistic effect  plays a considerable role, only if the relative velocity of the collision is over 30$\%$ of the speed of light, and it roughly corresponds to the collision energy 30 MeV per nucleon.

Ground states and some excited states of stable nuclei (in the following, self-bound nuclei) are classified to the localized self-bound system.
Each self-bound system is the solitary wave in the soliton theory, because it is satisfied that
\begin{itemize}
\item a self-bound nucleus moves stably without changing the shape
\end{itemize}
if there is no collision between nuclei.
Therefore the existence of solitary wave is trivially true for many-nucleon systems, where this issue is ultimately examined by the nonlinear framework with the ultimately and uniquely determined density-functional (i.e. self-consistent framework).
In other words, all the self-bound nuclei are the candidate of soliton.
All we have to do for verifying the soliton existence is to check 
\begin{itemize}
\item ${\rm [conditional]}$ a nucleus moves stably by preserving momentum and shape even after the collisions; 
\item the nucleus possibly experiences the phase shift and the time delay.
\end{itemize}
The first condition is expected to be satisfied conditionally. 
On the other hand, the second condition is trivially satisfied in case of atomic nuclei, as phase shifts have been observed and theoretically calculated in nuclear reactions, as well as the time delay.
One of the general motivation is to find a valid condition for the existence of nuclear soliton.

\subsection{Theoretical framework}
Among several theoretical models in nuclear physics, nuclear time-dependent density functional theory [Di30, En75] (TDDFT, for short), which describes the nuclear collision dynamics with the nucleon degree of freedom, is a unique theory including  time dependence, nonlinearity and the dispersive property simultaneously.
The solution of the TDDFT shows the unitary time evolution, which is preferable because of exact conservation of the total energy.
The dispersive property is satisfied by the non-relativistic theory, while it is violated in the massless relativistic theories.
In this context we remind the sine-Gordon equation is known as a soliton equation.
Furthermore it is worth noting here that, among sub-atomic theories except for the TDDFT, it is not easy to find a calculationally-feasible theoretical framework including the time-dependence.
Note that the TDDFT is also called nuclear time-dependent Hartree-Fock theory, and nuclear reaction is often referred to heavy-ion collision or ion collision.  
The theory with nucleon degree of freedom is called the microscopic treatment, because nucleus is a smaller component building up a nucleus.
The TDDFT is usually calculated in three-dimensional space, and the TDDFT have many stable localized stationary solution corresponding to the self-bound nuclei.
Nonlinearity, dispersive property, and the unitary time-dependence realized in the TDDFT are preferable for examining the soliton existence. 
Furthermore, nuclear saturation property brings about rather universal shallow potential well with the depth 50 MeV at the deepest, whose environmental setting is ideal to the existence of certain kinds of shallow water wave.

Before moving on to the nuclear theoretical models, a few remarks are made on the multi-dimensional treatment.
Quite limited things are known for the multi-dimensional soliton, where the shape of colliding waves play more roles.
In multi-dimensional case the soliton existence depends on whether the waves are spatially finite or not, and whether the waves are spherical or deformed.
As a multi-dimensional version of KdV equation, Kadomtsev-Petviashvili equation (KP equation, for short) is known.
In particular multi-dimensional version of NLS equation (\ref{eq:02}) cannot have the self-bound solution, while the multi-dimensional NLS type equation
\begin{equation} \begin{array}{ll}
\displaystyle  i \partial_t u +  \partial_x^2 u + \partial_y^2 u + k |u|^2 u = u \partial_x v, \vspace{1.5mm} \\
\partial_x^2 v - \partial_y^2 v  = -2 \partial_x (|u|^2)
\label{eq:022}
\end{array} \end{equation}
is known to have the soliton (or dromion) solution instead [Da74, Bo89, Hi90], where $v$ means the velocity potential.
Roughly speaking, the addition of nonlinear term contributes to keep the soliton property in this case.

\subsubsection{One-dimensional soliton model}
Let us begin with reviewing the preceding work on soliton propagation in nuclear physics.
In one-dimensional space, the Hamiltonian of $N$ bosons interacting through a $\delta$-force is represented by
\begin{equation} \begin{array}{ll}
\displaystyle  H = -\frac{1}{2} \sum_{i=1}^{N} \partial_{x_i}^2 -  v \sum_{i<j =1}^{N} \delta (x_i - x_j).
\label{eq14}
\end{array} \end{equation}
The corresponding stationary and non-stationary problems are known to be exactly solvable for bound states and for scattering states [Be3, Do76, Mc31, Ya67, Yo77, Za72].
Application of the variational principle to
\begin{equation} \begin{array}{ll}
\displaystyle <\Psi | \partial_t  - H  | \Psi > = N \int dx 
\left(   \psi^* i \partial_t  \psi  +  \frac{1}{2}\psi^*  \partial_x^2  \psi  + \frac{v}{2}(N-1)  \psi^*  \psi^*  \psi  \psi  \right)
\label{eq15}
\end{array} \end{equation}
leads to
\begin{equation} \begin{array}{ll}
 i \partial_t  \psi  +  \frac{1}{2} \partial_x^2  \psi  + \frac{v}{2}(N-1)   | \psi|^2  \psi = 0,
\label{eq16}
\end{array} \end{equation}
where $\Psi$ means the many-nucleon wave function and $\psi$ denotes single-nucleon wave function.
The similarity to NLS equation (\ref{eq:02}) is clear, so that the soliton solution follows.
The static solution is
\begin{equation} \begin{array}{ll}
\psi_i (x) = \frac{\sqrt{(N-1) v}}{2 {{\rm cosh}((N-1)vx/2)}}
\label{eq17}
\end{array} \end{equation}
with the energy
\begin{equation} \begin{array}{ll}
E_H = - \frac{N(N-1)^2 v^2}{24}
\label{eq18}
\end{array} \end{equation}
and the density
\begin{equation} \begin{array}{ll}
 \rho(x) =  \frac{N(N-1) v} {4 {{\rm cosh}^2((N-1)vx/2)}}.
\label{eq19}
\end{array} \end{equation}
For $2N$ particle case, two-soliton solution is obtained.
The two-soliton solution is represented by
\begin{equation} \begin{array}{ll}
\psi (t,x) = 
\frac{\sqrt{2(N-1) v}}{2}
e^{-(i/2)(K^2-a^2)t} \vspace{1.5mm} \\ \qquad
\frac{ e^{iKx} \{ e^{-a(x-Kt)} + (K^2/(K-ia)^2) e^{-a(3x+Kt)} \} + (K \leftrightarrow -K) }
  {1 + 2e^{-2ax} {\rm cosh}(2aKt) - 2a^2e^{-2ax} {\rm Re} (e^{2iKx}/(K+ia)^2) + (K^4/(K^2+a^2)^2) e^{-4ax} }.
\label{eq20}
\end{array} \end{equation}
The existence of two-soliton solution ensures the existence soliton in a given theoretical framework.

\subsubsection{Three-dimensional model}
Two-dimensional model is realized as the axial symmetric model in nuclear density functional theory dealing with finite quantum systems, and the axis of symmetry is taken as the collision axis in the time-dependent collision calculations.
In this sense two dimensional calculation computes one-dimensional colliding motion along the center axis.
One and two-dimensional models are toy models for simulating the collision, because the effect described by the outer product (vector product) cannot be rigorously incorporated.
Consequently spin effect on the dynamics such as spin-orbit force effect cannot be rigorously treated in one and two dimensional models (cf. the representation of spin current ${\bm J}({\bm r})$ in Eq.~(\ref{eq21-2})).
Note that spin-orbit force in the non-relativistic framework arises from the special relativity theory.
In particular the spin orbit force has well known to play a decisive role in the structure of nuclei (cf. magic numbers of nuclear structure [Ri80]).

Let us consider three-dimensional case.
It is remarkable that nuclear medium as nucleon degree of freedom consists of two different kinds of fermions: protons and neutrons.
In the following the formalism of TDDFT [Di30, En75] for low-energy nuclear reactions are introduced based on [Bo05], where the Skyrme interaction [Sk56] is utilized as the effective nuclear force in the most of TDDFT calculations.
The Skyrme interaction is a zero-range formalism of effective nucleon-nucleon interaction. 
The TDDFT with Skyrme type zero-range interaction is represented by several densities
\begin{equation} \begin{array}{ll}
\displaystyle \rho({\bm r}) = \sum_{i,\sigma}(\psi_i^* ({\bm r},\sigma)  \psi_i({\bm r},\sigma)),   
\qquad \displaystyle \tau({\bm r}) = \sum_{i,\sigma} (\nabla \psi_i^* ({\bm r},\sigma) \cdot \nabla \psi_i({\bm r},\sigma) ),   \vspace{1.5mm} \\
\displaystyle {\bm j}({\bm r}) = \frac{1}{2i} \sum_{i,\sigma} (\psi_i^* \nabla \psi_i({\bm r},\sigma) - \psi_i \nabla \psi_i^*({\bm r},\sigma) ),  
\label{eq21-1}
\end{array} \end{equation}
and
\begin{equation} \begin{array}{ll}
\displaystyle {\bm s}({\bm r}) =  \sum_{i,\sigma,\sigma'} (\psi_i^* ({\bm r},\sigma)  \psi_i({\bm r},\sigma') \langle \sigma | {\hat \sigma} | \sigma' \rangle ),   
\qquad \displaystyle {\bm T}({\bm r}) =  \sum_{i,\sigma} (\nabla \psi_i^* ({\bm r},\sigma) \cdot \nabla \psi_i({\bm r},\sigma') \langle \sigma | {\hat \sigma} | \sigma' \rangle ),   \vspace{1.5mm} \\
\displaystyle {\bm J}({\bm r}) =  \frac{1}{2i} \sum_{i,\sigma} (\psi_i^* \nabla \psi_i({\bm r},\sigma) - \psi_i \nabla \psi_i^*({\bm r},\sigma) ) \times \langle \sigma | {\hat \sigma} | \sigma' \rangle ), 
\label{eq21-2}
\end{array} \end{equation}
where $ \psi_i({\bm r},\sigma)$ and $ \psi_i^*({\bm r},\sigma)$ are $i$-th single wave function and its complex conjugate respectively, $\rho({\bm r})$, $\tau({\bm r})$ and ${\bm j}({\bm r})$ denote the density, the kinetic energy density and the momentum density respectively, and ${\bm s}({\bm r})$, ${\bm T}({\bm r}) $ and ${\bm J}({\bm r})$ stand for the spin density, the spin kinetic density and the spin current density respectively. 
Single wave functions depend on both spatial variable ${\bm r} \in {\mathbf R}^3$ and the spin $\sigma$, while the spin dependence is summed up in each density. 
By assuming wave functions and densities as depending also on the time variable $t \in {\mathbf R}$, each single-nucleon satisfies the equation of the form.
\begin{equation} \begin{array}{ll}
\displaystyle i \hbar \partial_t \psi_i (t,{\bm r},\sigma) =   h \psi_i(t,{\bm r},\sigma) 
\label{eq22}
\end{array} \end{equation}
with 
\begin{equation} \begin{array}{ll}
\displaystyle h \psi_i(t,{\bm r},\sigma) = \sum_{\sigma'} \Bigg[
- \nabla \cdot \frac{\hbar^2}{2 m^*_q} \nabla \delta_{\sigma,\sigma'} 
+ U_q({\bm r})  \delta_{\sigma,\sigma'} 
 + {\bm V}_q({\bm r}) \cdot \langle \sigma | {\hat \sigma} | \sigma' \rangle
  + i C_q({\bm r}) \cdot \nabla \delta_{\sigma,\sigma'} \vspace{1.5mm} \\ \qquad
  + i {\bm W}_q({\bm r}) \cdot (  \langle \sigma | {\hat \sigma} | \sigma' \rangle \times \nabla )
\Bigg]  \psi_i(t,{\bm r},\sigma') ,
\label{eq23}
\end{array} \end{equation}
where $h$ is the single-particle Hamiltonian, $m^*_q$ denotes the effective mass, and $U_q$, ${\bm V}_q$, $C_q$, and ${\bm W}_q$ mean the spin scalar potential, the spin vector potential, the current potential, and the spin orbit potential, respectively. 
The isospin index $q$ distinguishes protons ($q = p$) from neutrons ($q = n$).
For realizing of fermionic statistical property, single wave functions are assumed to form the single Slater determinant, where this assumption is necessary to derive Eq.~(\ref{eq23}).
First of all, the nucleon-nucleon interaction is fully represented by the densities, and here is the reason why this formalism is called the nuclear TDDFT.
In the second this formalism tells us that each single nucleon does not interact directly with the other nucleon, but with the force field described by the collectively summed-up densities (\ref{eq21-1}) and  (\ref{eq21-2}).
Here is the reason why the nuclear TDDFT is claimed to be the theory based on the mean-field description of many-body interaction (in the same context, the nuclear TDDFT is also called the nuclear TDHF).
Furthermore one-body dissipation with the unitarity appears mainly due to the internal excitation of nucleus.
Note that the concept of one-body dissipation is a kind of dissipation, but it does not violate the unitarity of time evolution.  
The details are given by
\begin{equation} \begin{array}{ll}
 \frac{\hbar^2}{2m_q^*} = \frac{\hbar^2}{2m_q} + B_3 \rho + B_4 \rho_q   \vspace{3.5mm} \\
 U_q({\bm r})  = 2 B_1 \rho  +  2 B_2 \rho_q +  B_3 (\tau + i \nabla \cdot {\bm j}) 
 + B_4  (\tau_q + i \nabla \cdot {\bm j}_q)   \vspace{1.5mm}  \\ \qquad
 + 2 B_5 \triangle \rho + 2 B_2 \triangle \rho_q + (2 + \alpha) B_7 \rho^{\alpha+1}  \vspace{1.5mm}  \\ \qquad
 + B_8 \left\{ \alpha \rho^{\alpha-1} (\rho_n^2 + \rho_p^2) + 2 \rho^{\alpha} \rho_q 
 + B_9 (\nabla \cdot {\bm J} + \nabla \cdot {\bm J}_q) \right\}   \\ \qquad
 + \alpha \rho^{\alpha-1} \{ B_{12} {\bm s}^2 + B_{13} {\bm s}_n^2 ({\bm s}_n^2 + {\bm s}_p^2  )   \}
 + \left[ e^2 \int \frac{\rho_p({\bm r}')}{|{\bm r} - {\bm r}'|} d{\bm r}'  -e^2 \left( \frac{3 \rho_p}{\pi} \right)^{1/3} \right]  \delta_{q, p} \vspace{2.5mm} \\
  {\bm V}_q({\bm r}) = B_9 (\nabla \times {\bm j} + \nabla \times {\bm j}_q)  + 2 B_{10} {\bm s} + 3 B_{11} {\bm s}_q  \vspace{1.5mm} \\ \qquad
 + 2 \rho^{\alpha} (2 B_{12} {\bm s} + 2 B_{13} {\bm s}_q) + B_9 (\nabla \times {\bm J} + \nabla \times {\bm J}_q)   \vspace{3.5mm} \\
 {\bm C}_q({\bm r})  = 2 B_3 {\bm j} + 2 B_4 {\bm j}_q - B_9 (\nabla \times {\bm s}+\nabla \times {\bm s}_q) \vspace{3.5mm} \\
 {\bm W}_q({\bm r}) = - B_9 (\nabla \rho + \nabla \rho_q) ,
\label{eq24}
\end{array} \end{equation}
where a part shown inside the parenthesis $\left[ \cdot \right]$ in $U_q({\bm r})$ shows the Coulomb interaction acting only on protons.
13 different coefficients ($B_1, B_2, \cdots, B_{13}$) must be determined, while they are reduced to only 10 parameters ($t_0, t_1, \cdots, x_3, \alpha$).
For the derivation of the above effective nuclear interaction, see [Va72, Bo05].

\begin{table}
\caption{Parameter setting in the TDDFT. The reduced coefficients (a) and a Skyrme parameter set (b) are shown.
Among many parameter sets (models for the effective nuclear force), the SV-bas model is taken in this paper. \\ }
\begin{center} \begin{tabular}{l}
(a) Reduced coefficients   \vspace{0.6mm} \\ \hline \vspace{-0.9mm} \\
 $ B_1 = t_0 (1+x_0/2)/2$ \vspace{1.5mm} \\
 $B_2 = -t_0 (x_0+1/2)/2$ \vspace{1.5mm} \\
  $B_3 =  (t_1 + t_2)/4 +  (t_1 x_1 + t_2 x_2)/8$ \vspace{1.5mm} \\
  $B_4 =  (t_2 - t_1)/8 -  (t_2 x_2 - t_1 x_1)/4$ \vspace{1.5mm} \\
  $B_5 =  (t_2 - 3t_1)/16 +  (t_2 x_2 - 3 t_1 x_1)/32$ \vspace{1.5mm} \\
  $B_6 =  (3t_1 + t_2)/32 +  ( t_1 x_1 + t_2 x_2)/16$ \vspace{1.5mm} \\
  $B_7 =  t_3 (1 + x_3/2)/12$ \vspace{1.5mm} \\
  $B_8 =  - t_3 (x_3 + 1/2)/12$ \vspace{1.5mm} \\
  $B_9 =  - W_0/2$ \vspace{1.5mm} \\
  $B_{10} =  - t_0 x_0/4$   \vspace{1.5mm} \\
  $B_{11} =  - t_0/4$ \vspace{1.5mm}  \\
  $B_{12} =  -t_3 x_3/24$ \vspace{1.5mm} \\
  $B_{13} =  - t_3/24$ \vspace{1.5mm} \\
  \hline 
\label{table1}
\end{tabular}
\hspace{18mm}
\begin{tabular}{l}
(b) Skyrme parameter set (SV-bas) [Kl09]  \vspace{0.6mm} \\ \hline \vspace{-0.9mm} \\
  $t_0 = -1879.640018 ~[ {\rm MeV \cdot fm^3}]$  \vspace{1.5mm} \\
  $t_1 = 313.7493427 ~ [{\rm MeV \cdot fm^5}] $  \vspace{1.5mm} \\
   $t_2 = 112.6762700 ~ [{\rm MeV \cdot fm^5}] $ \vspace{1.5mm} \\ 
   $t_3  = 12527.38921 ~ [ {\rm MeV \cdot fm^{3+3 \alpha} }] $  \vspace{1.5mm} \\ 
   $W_0 = 124.6333000 ~ [ {\rm MeV \cdot fm^5}]$ \vspace{1.5mm} \\ 
    $x_0 = 0.2585452462$  \vspace{1.5mm} \\ 
    $x_1 = -0.3816889952$ \vspace{1.5mm} \\  
   $x_2 = -2.823640993$  \vspace{1.5mm} \\
  $x_3 = 0.1232283530$  \vspace{1.5mm} \\
     $\alpha = 0.3$  \vspace{1.5mm} \\
  \hline  \vspace{17mm}
\label{table2}
\end{tabular}
 \end{center}
\end{table}

Although more than 100 parameter sets are proposed for the Skyrme-type effective nuclear interaction (the values for $\{ t_0, t_1, t_2, t_3, W_0, x_0, x_1, x_2, x_3 \}$), the ultimate parameter set has not been known including such an existence.
Here we take SV-bas parameter set (Table~\ref{table1}).
The SV-bas parameter set is known for well reproducing the neutron skin thickness of heavy nuclei such as $^{208}$Pb (for a compilation of experimental and theoretical results,see [Ne15]).
The quality of SV-bas in some relevant heavy nuclei can be found in [Iw19].
On the other hand, the description of light ions (helium isotopes) using SV-bas is also confirmed to be sufficiently good [Iw15].
The pairing interaction is not introduced in the present density functional, as the collision energy of the present study is sufficient high for pairing interaction not to play a significant role. 
Indeed, from an energetic point of view, the nuclear pairing is the effect less than a few 100s of  keV per nucleon.
A set of equations (\ref{eq22}), (\ref{eq23}) and (\ref{eq24}) are called the nuclear TDDFT or the nuclear TDHF equations.
The nuclear TDDFT is known to reproduce the result rather sufficiently nowadays (for recent reviews, see [Si18, St19, Se19]).

\subsection{Solitons in many-nucleon systems}

\subsubsection{Similarity of master equations}
For verifying the soliton existence, we begin with finding the similarity between NLS (\ref{eq04}) and the nuclear TDDFT.
For one-dimensional cases, as the soliton solution has been obtained, Eq.~(\ref{eq16}) is essentially identical with Eq.~(\ref{eq09}). 
For three-dimensional cases, a term with $\hbar^2/2m_q^*$ in the TDDFT corresponds to the second term of the left hand side of Eq.~(\ref{eq09}). 
Here we see that the TDDFT is a  Schr\"{o}dinger type equation.
Meanwhile the nonlinear term $|u|^2 u$ in Eq.~(\ref{eq09}) corresponds to terms with the coefficients $B_1$ and $B_2$ (depending essentially on the parameter $t_0$).
The terms with the coefficients $B_7$ and $B_8$  (depending essentially on the parameter $t_3$), which are known to be indispensable to reproduce the nuclear saturation properties [Va72], are also relevant, because they introduce additional fractional power contributions (cf. $\alpha$ in the Skyrme parameter set: the fractional power).
The dominance of $t_0$ and $t_3$ terms has been confirmed for the binding energies of $^4$He and $^{8}$He [Iw15], where the experimental binding energy is 28.30 MeV and 31.40 for $^4$He and $^{8}$He respectively.
No self-bound states of $^4$He and $^{8}$He are obtained if $t_0$ term is turned off, and even in the presence of $t_0$ term the calculated binding energies are at the order of 1000 MeV which are far from the realistic binding energy.
More quantitatively, with respect to the bindings of $^{4}$He, a large binding (due to the attractive force property of $t_0$ term) at the order of 1000 MeV is obtained only by $t_0$ term, it is substantially modified by $t_3$ term (due to the repulsive force property of $t_3$ term) as 63.80 MeV, and the momentum density contribution ($t_1$ term) reduces it to a realistic value 27.71 MeV, where the binding energy calculated by including all the terms is  27.73 MeV.
Note that the spin-orbit contribution is known to be important in the nuclear structure, but it does not play a prominent role in this case because $^{4}$He is a spin-saturated system.
Rough estimation tells us that interaction part of the TDDFT with SV-bas model (the inhomogeneous term of nonlinear Schr\"{o}dinger type equation) is dominated by the $t_0$ and $t_3$ terms with the percentage:
\[ \frac{|63.80|}{|63.80| + |63.80-27.71| + |27.71-27.73|} \times 100 = 63.9 \%,  \]
where the amplitudes of $t_0$ and $t_3$ terms, $t_1$ term and the other terms are estimated as  $|63.80|$, $|63.80-27.71|$, and $|27.71-27.73|$, respectively.
Dominance of those terms in the nuclear density functional implies the validity of an energy-dependent soliton existence in which  $t_0$ and $t_3$ terms are responsible for the soliton existence and energy dependence respectively.
This similarity between NLS and the nuclear TDDFT provides us a sound motivation to investigate the soliton propagation in nuclear TDDFT.

\begin{table}
\caption{Self-binding energies of the ground states [Iw19] are calculated using the SV-bas effective nuclear force.
Binding energy per nucleon (MeV) of the initial nuclei are compared to those of intermediate fused system, and the corresponding experimental values are shown in parenthesis [Nud]. \\ }
\begin{center} \begin{tabular}{rccc}
$(A,Z)$ \quad &B($^AZ$)  \quad  & B($^{A+4}Z$)  \quad & B($^{2A+4}2Z$)    \vspace{0.6mm} \\ \hline \vspace{-0.9mm} \\
(4,2)  & 6.93(7.08)  & 4.48(3.93) & 6.07(7.06)  \vspace{1.5mm} \\
(16,8) & 8.21(7.98) &  7.84(7.57) & 8.75(8.58) \vspace{1.5mm} \\
  \hline 
\label{table3}
\end{tabular}
 \end{center}
\end{table}


\begin{figure}[tb]
\begin{center}
\includegraphics[width=10cm]{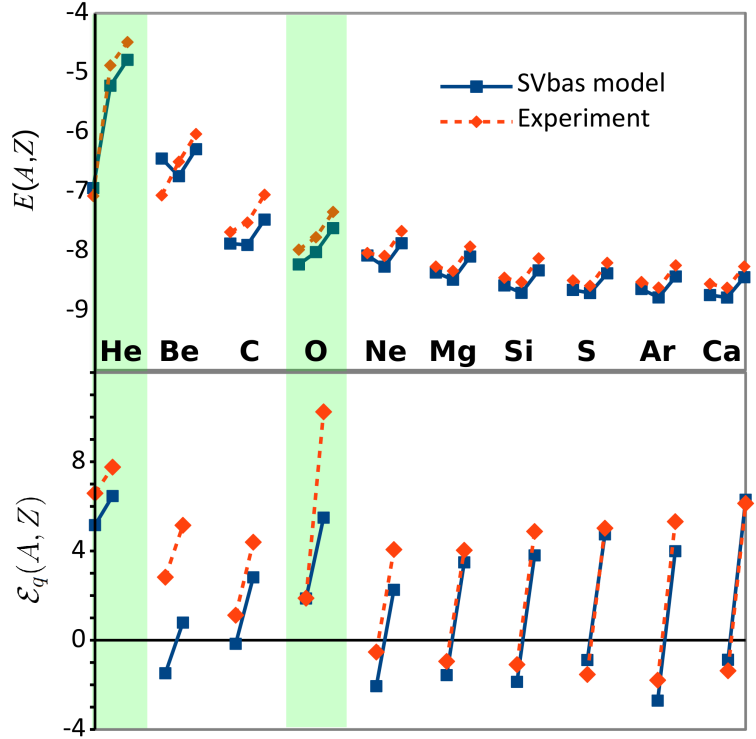}
\end{center} \label{fig:3}
\caption{(Color online) Single particle energy and the corresponding energy differences are compared to experiments [Nud]. 
[Upper panel] For $Z \le 20$ nuclei with $(A,Z)$=(4,2), (8,4), $\cdots$ (40,20), the binding energies $E(A,Z)$, $E(A+2,Z)$, $E(A+2,Z+2)$ are shown by the connected lines in this order in each column.
 [Lower panel] The corresponding energy difference ${\mathcal E}_n(A,Z)$ and ${\mathcal E}_p(A,Z) $ for each nucleus $^{Z}A$ is shown by the connected lines in this order.}
\end{figure}

\subsubsection{Mechanism of soliton propagation in the TDDFT}
Some specific physics associated with many-nucleon systems are presented with respect to the soliton propagation. 
In three-dimensional nuclear TDDFT, the existence of solitary wave corresponds to the existence of self-bound stationary states.
For low-energy nuclear reactions, fusion, deep inelastic collision, and collision-fission such as fusion-fission and quasi-fission are possible to appear.
Particularly, in case of fusion, the solitary waves are totally destroyed.
It implies that solitary wave cannot necessarily be the soliton, and the soliton existence is inevitably conditional. 
Let us begin with the collision between $^{4}$He and $^{8}$He.
Following the usage of nuclear reaction representation, the fusion reaction realized by collision between two self-bound nucleus $^{4}$He (helium 4: 2 protons and 2 neutrons) and $^{8}$He (helium 8: 2 protons and 6 neutrons) is represented by
\begin{equation} \begin{array}{ll}
\displaystyle ^8{\rm He} + ^4{\rm He}  ~\to~  ^{12}{\rm Be},
\label{eq25}
\end{array} \end{equation}
where $^{12}$Be  (beryllium 12: 4 protons and 8 neutrons) is produced as a result of fusion reaction.
Fusion reaction is generally an exothermic or endothermic reaction according to the total binding energy difference between reactants and products, where a chemical element iron ($Z=26$) is the most stable element.
On the other hand, if  self-bound states $^{4}$He and $^{8}$He hold the soliton property,
\begin{equation} \begin{array}{ll}
\displaystyle ^8{\rm He} + ^4{\rm He}  ~\rightleftharpoons~   ^4{\rm He}  +  ^8{\rm He}
\label{eq26}
\end{array} \end{equation}
takes place in which the total energy is conserved before and after the collision.
In the context of reaction theory, the soliton property is included in a class of reactions with the time-reversal symmetry. 
The goal is to find the condition for the appearance of soliton events shown by Eq~(\ref{eq26}). 
The time reversal symmetry arises from the energy conservation, according to Noether’s theorem, and the total energy is strictly conserved by the nuclear TDDFT framework.
For each collision there are two controllable parameters: the relative velocity of collision (i.e., the collision energy) and the impact parameter of collision (usually denoted by $b$~fm).
The condition for soliton existence is expected to be written by these two control parameters (i.e. the initial condition).

\begin{figure}[tb]
\begin{center}
\includegraphics[width=2cm]{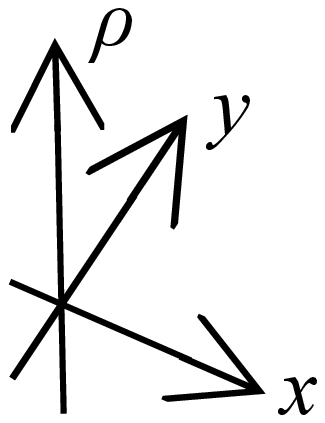}
\includegraphics[width=5cm]{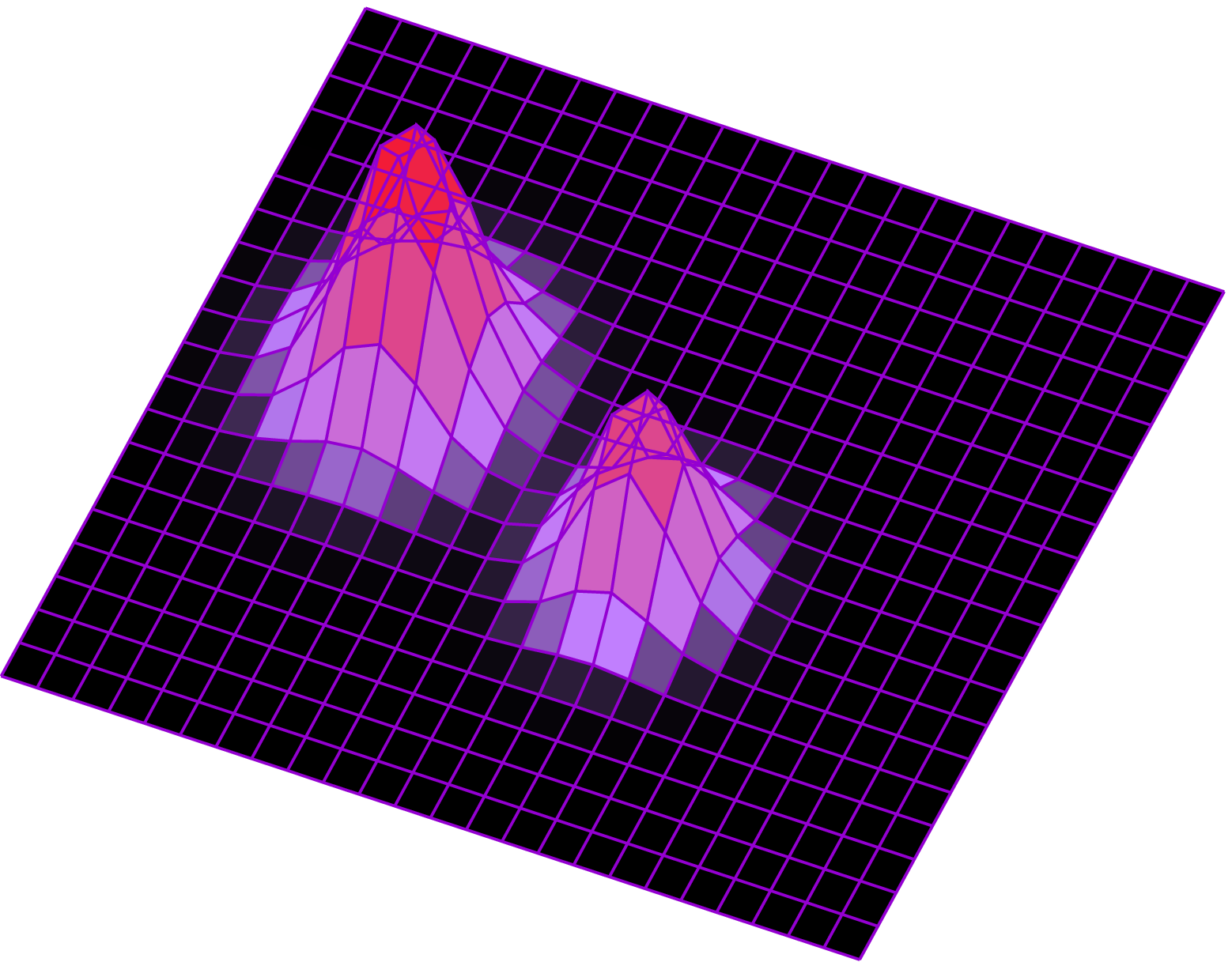}
\includegraphics[width=5cm]{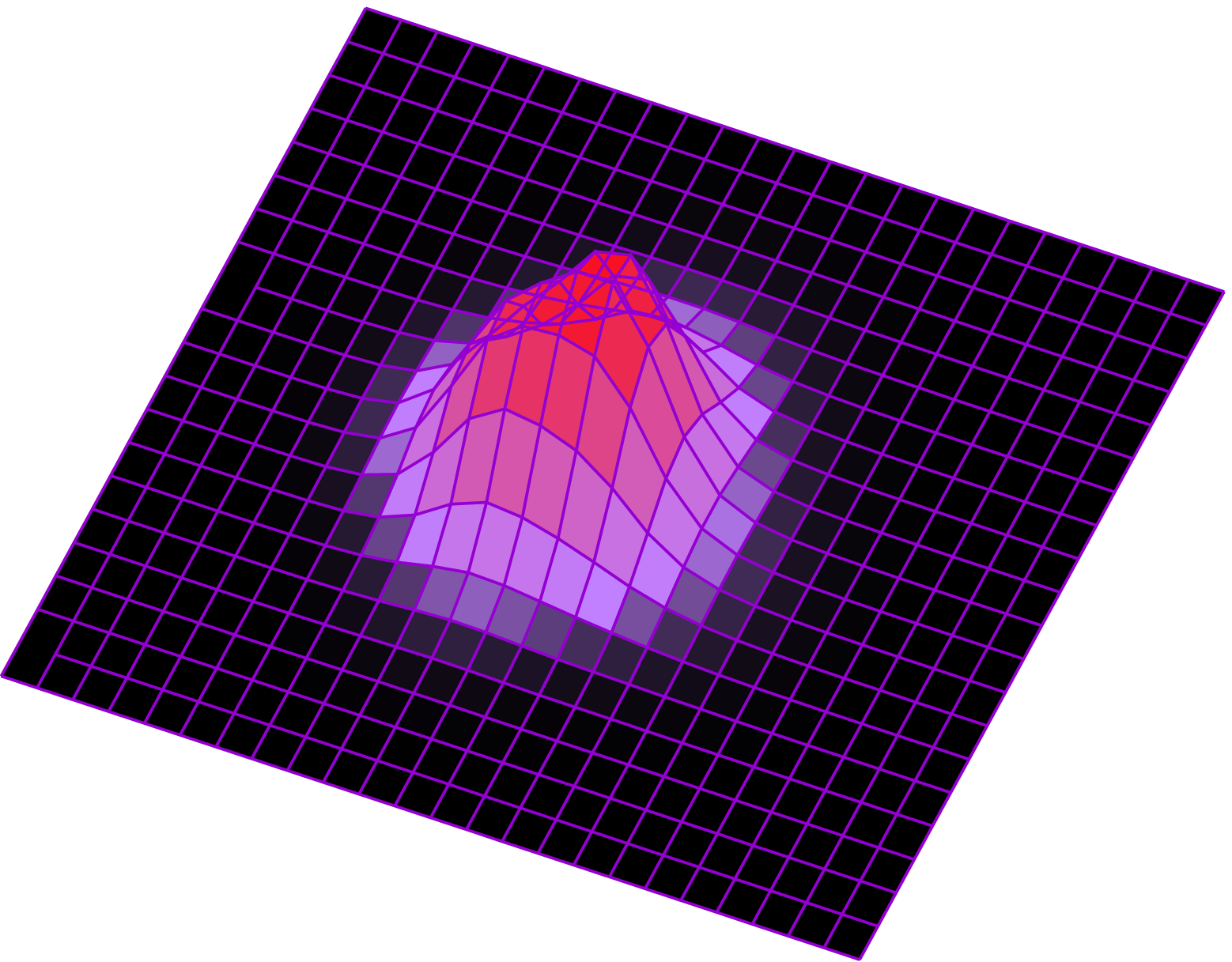}
\includegraphics[width=5cm]{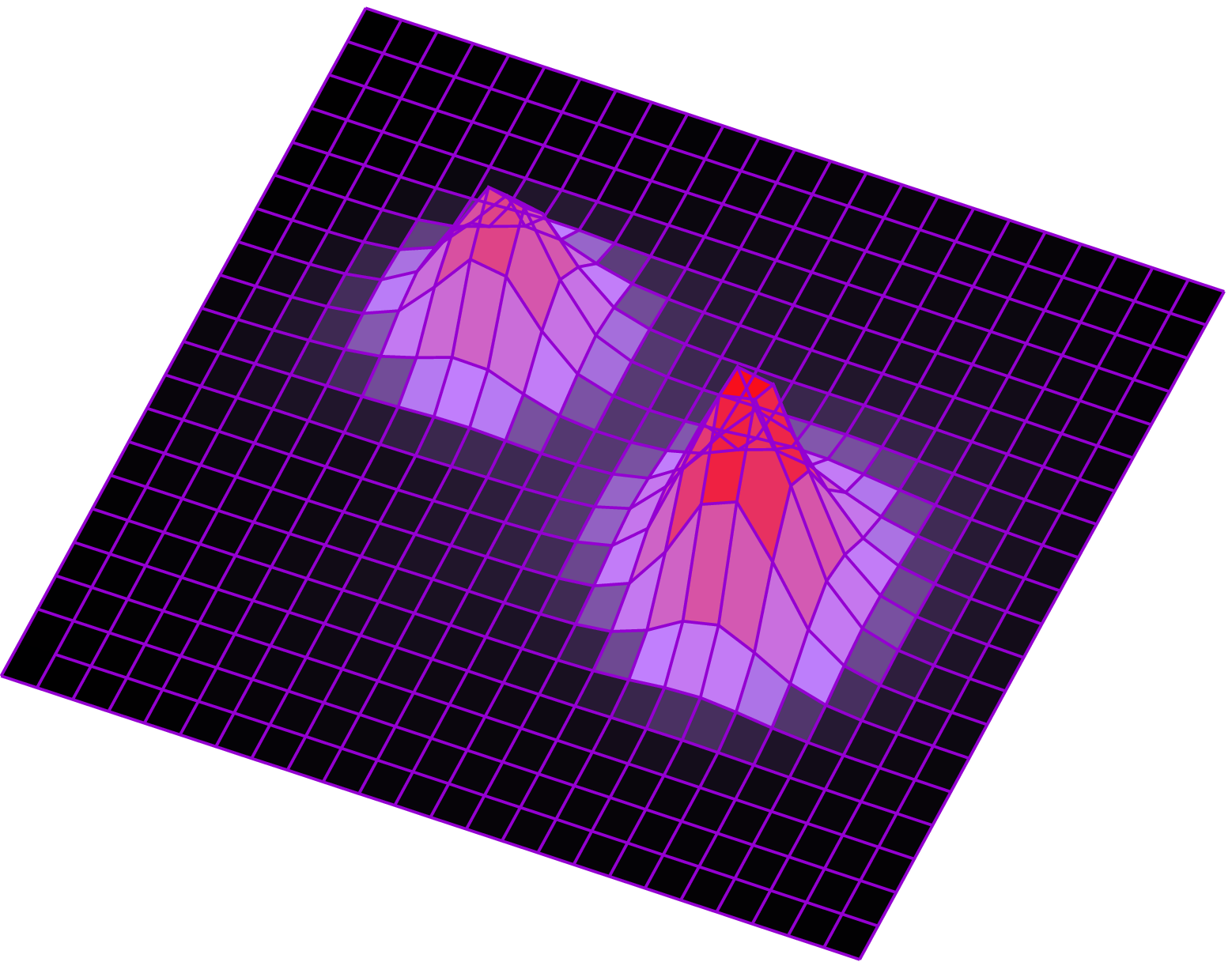} \vspace{1.5mm} \\
{\small \hspace{3cm}  $t$ = $8/3 \times 10^{-22}$s  \hspace{1.85cm} $t$ = $28/3 \times 10^{-22}$s  \hspace{1.85cm} $t$ = $48/3 \times 10^{-22}$s } 
\end{center}
\caption{(Color online) Imperfect soliton including the spin degree of freedom, fermionic statistical property, the multi-dimensionality, and the effect due to the non-central collision.
Time evolution of $^{8}$He ~+~ $^{4}$He for $E_K = 7.50 $~MeV and $b = 3.0$~fm are shown. 
The collision energy is around the upper-limit energy of fast charge equilibration.
That is, by increasing the energy, the transparent component becomes dominat.  
For better sights, time evolution of total density is depicted by projecting them on the reaction plane ($z = 0$).
The density is plotted on the vertical axis taken from 0 to 0.6 fm$^{-1}$, where the horizontal area is fixed to $(x,y) = 24 \times 20 $~fm$^2$.
In this situation, 0.29 protons are transferred from $^{4}$He to $^{8}$He, and 0.26 neutrons are transferred from $^{8}$He to $^{4}$He, where we can find a weak effect of dual-way type charge equilibration [Iw10n] leading to the contamination of pure soliton.
The self-bound property of $^{8}$He and $^{4}$He contributes to recover the original shape if the transparencies of both mass and momentum is sufficiently high (kinetic energy loss is less than 5 MeV (see Fig. 3)). }
 \label{fig:6}
\end{figure}

\begin{figure}[tb]
\begin{center}
\includegraphics[width=6cm]{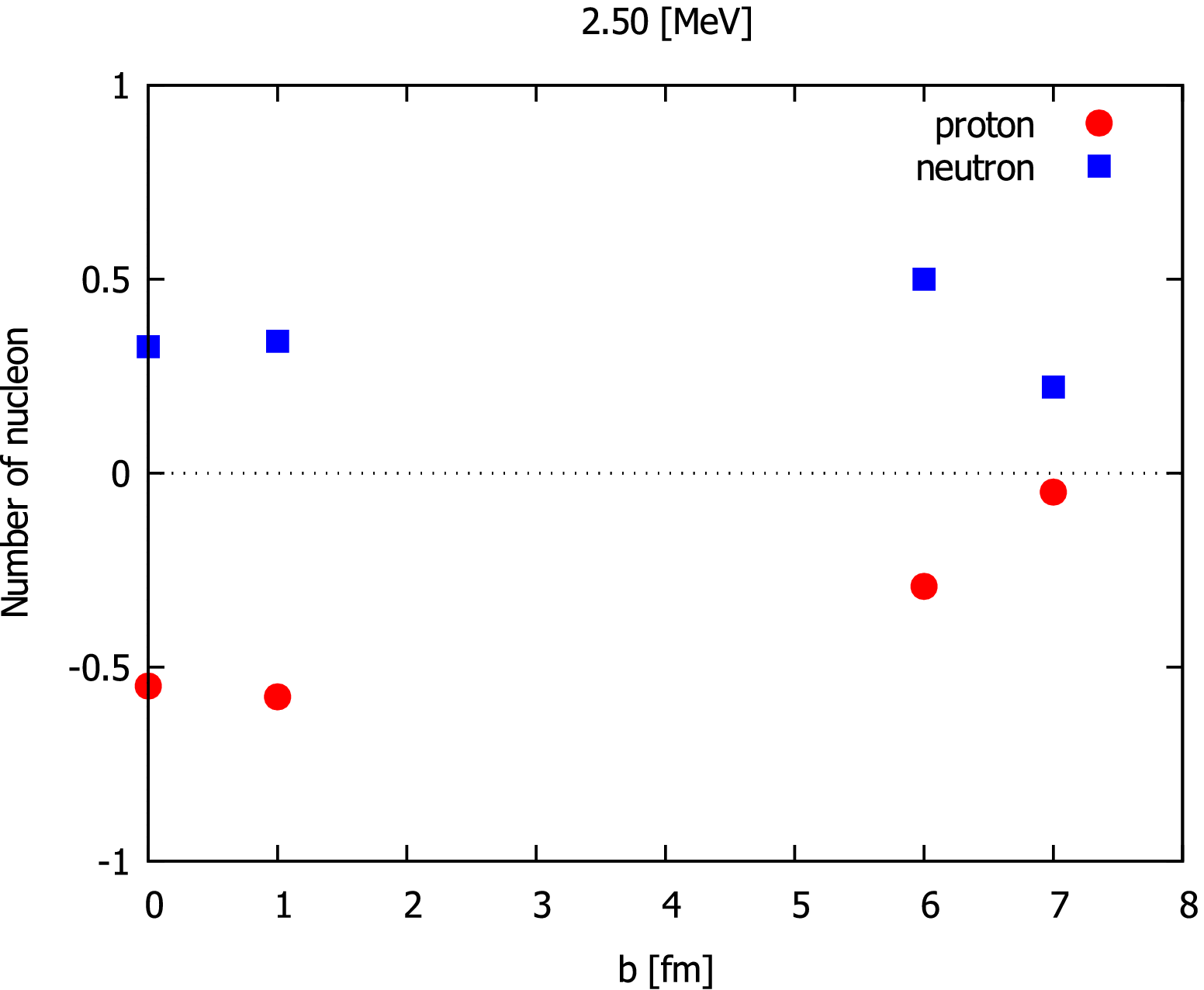} \qquad 
\includegraphics[width=6cm]{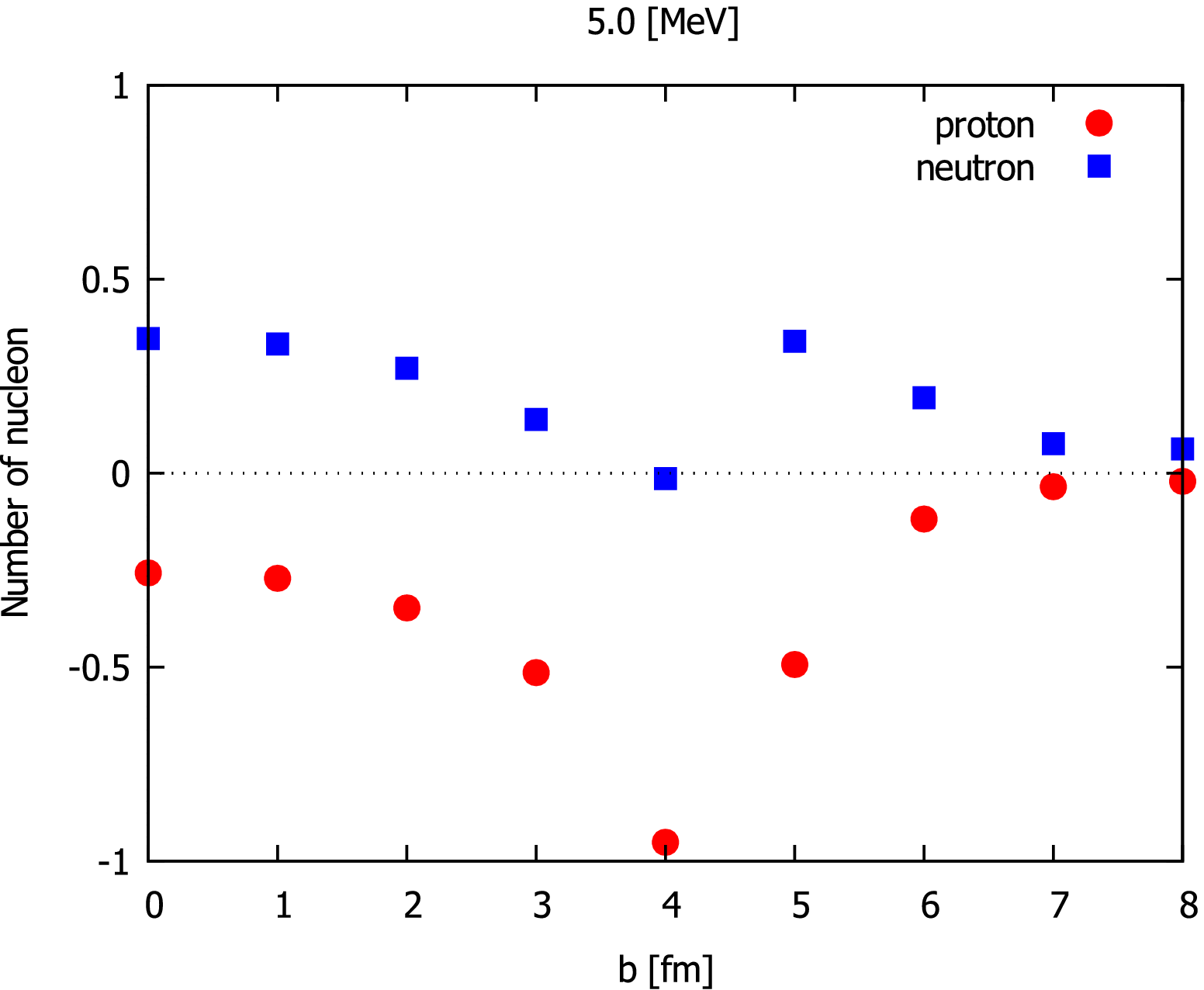} \vspace{5mm} \\ 
\includegraphics[width=6cm]{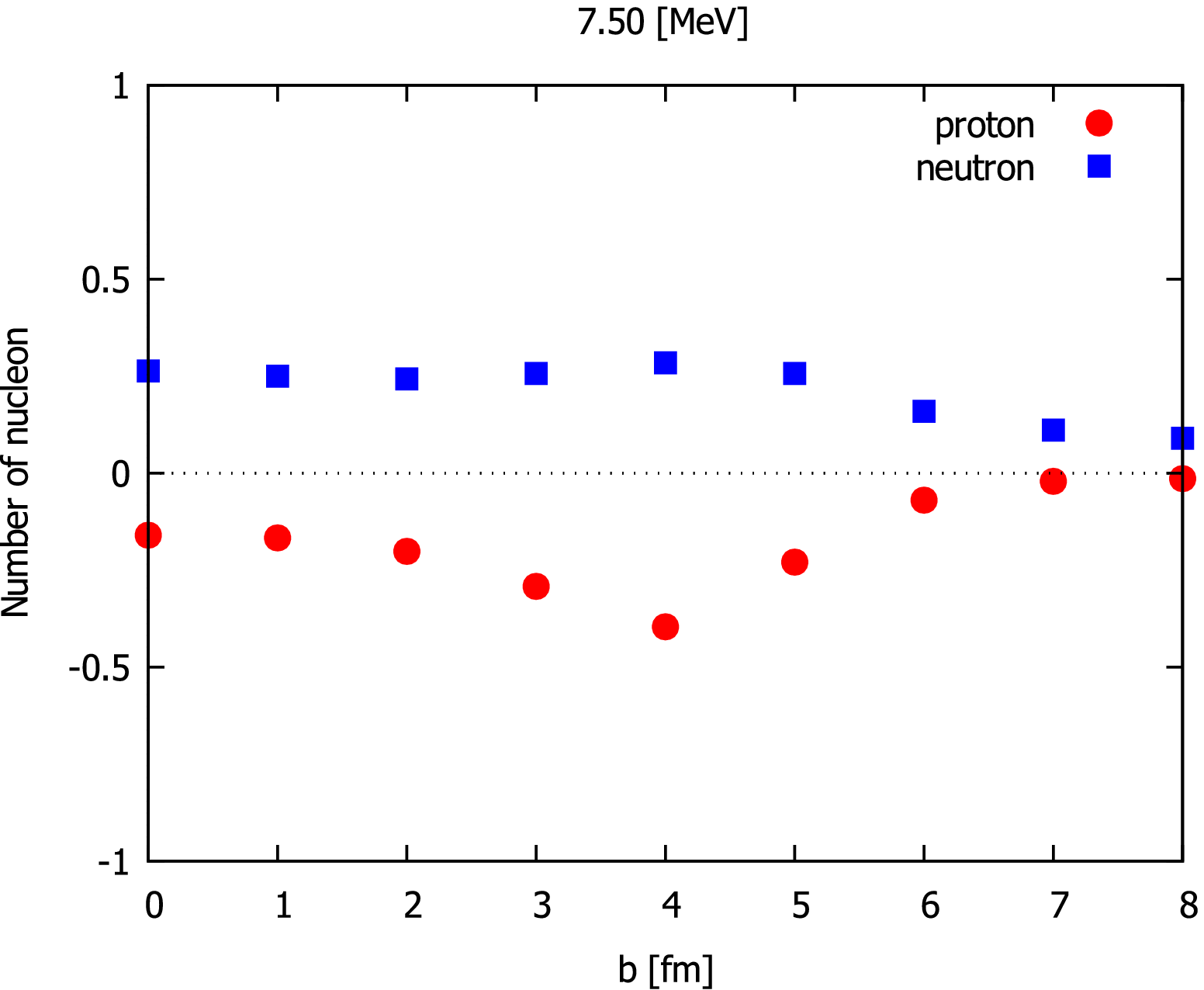} \qquad 
\includegraphics[width=6cm]{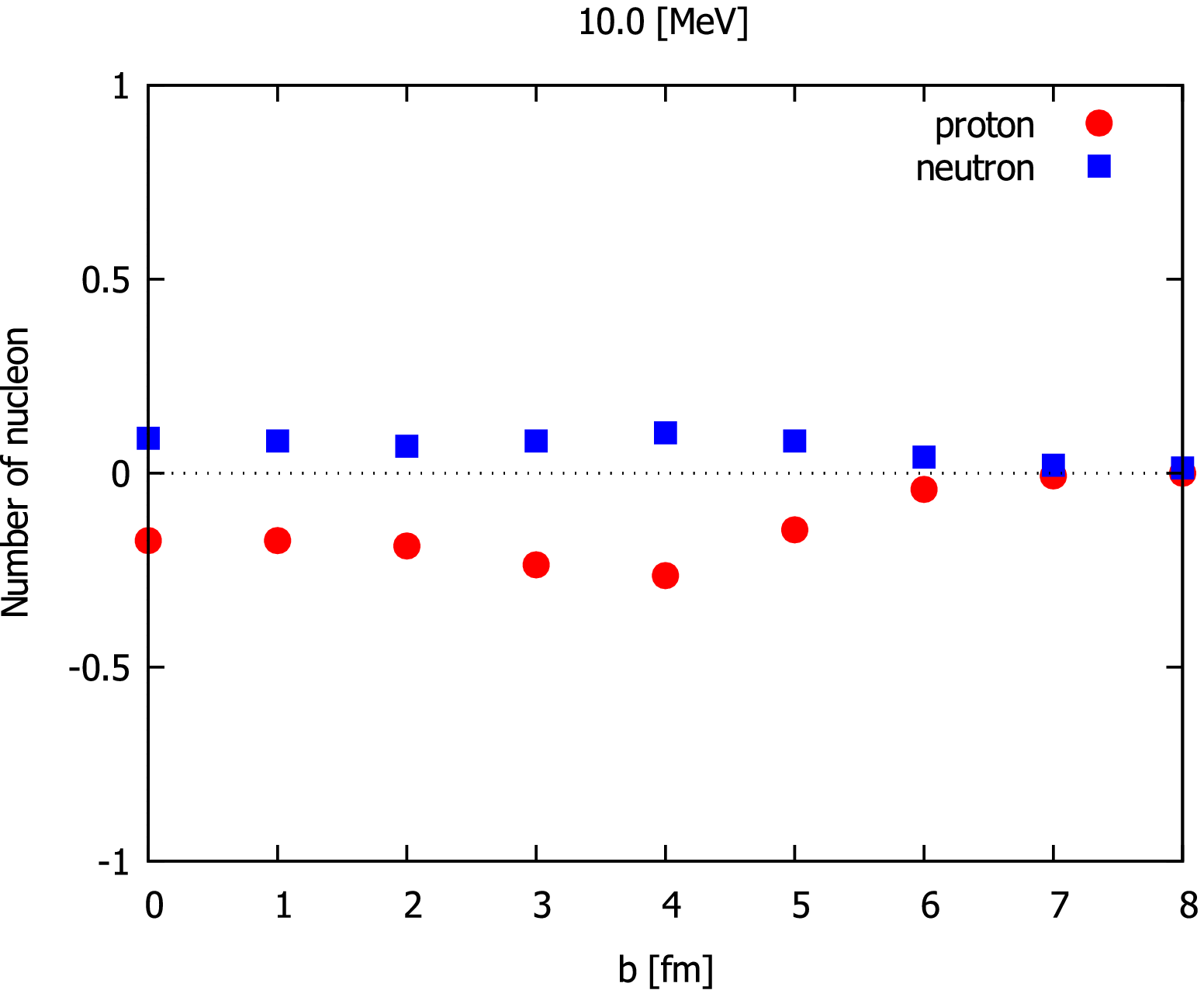} \vspace{5mm} \\ 
\includegraphics[width=6cm]{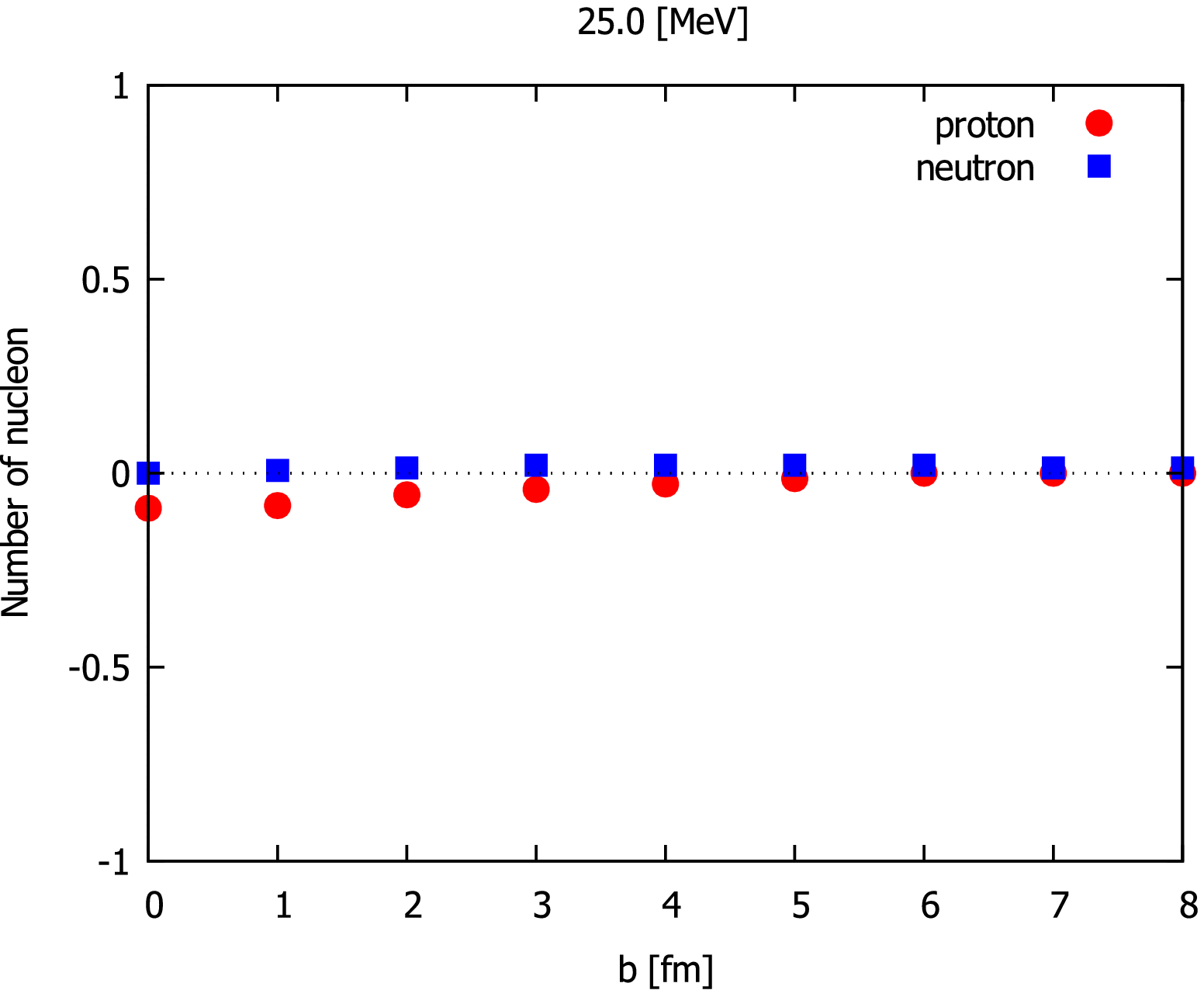} \qquad 
\includegraphics[width=6cm]{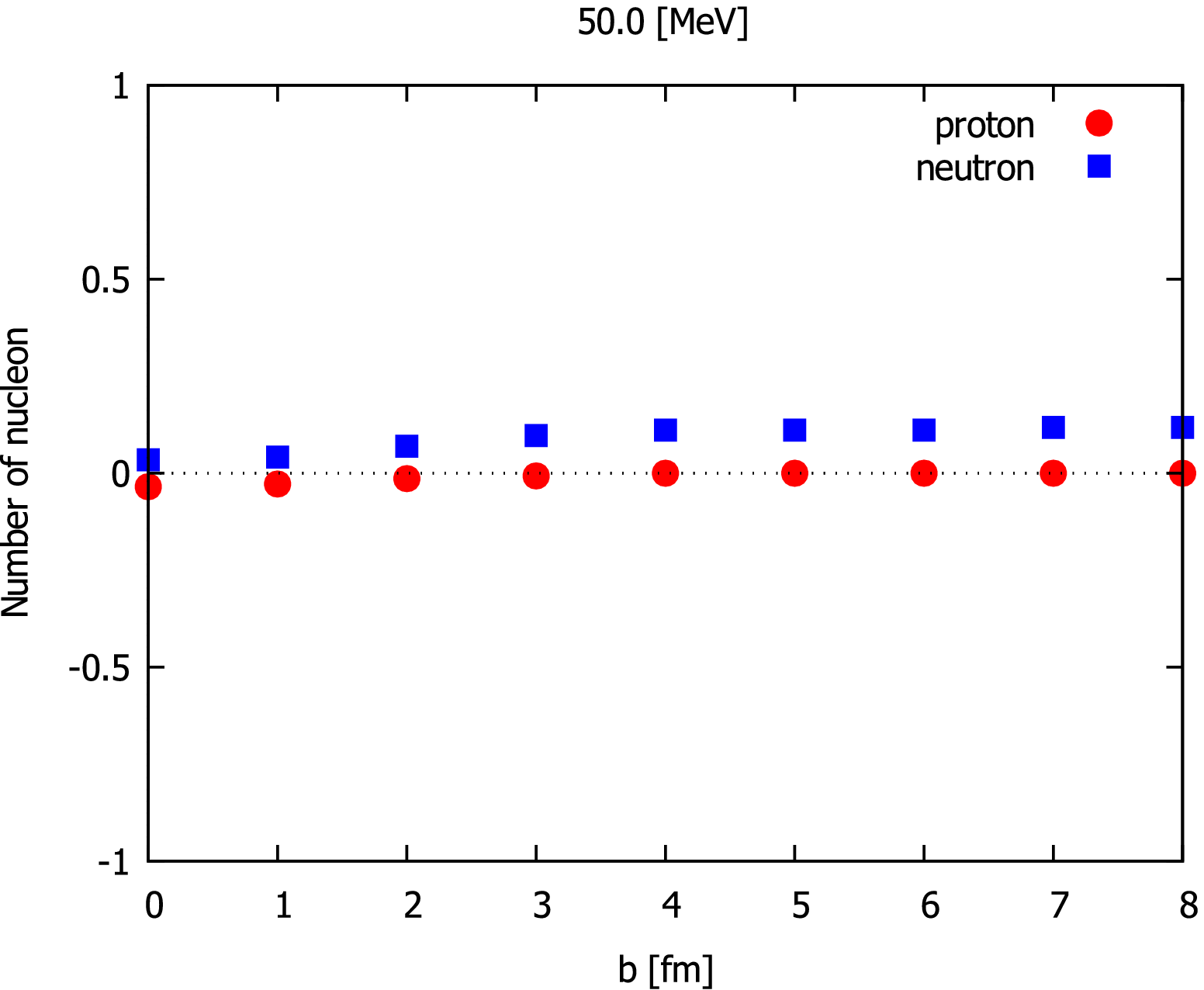} \\
\end{center}
\caption{(Color online) For collisions $^{8}$He ~+~ $^{4}$He, transferred nucleon from $^{8}$He to $^{4}$He are shown depending on the relative velocity of the collisions.
The impact parameter dependence with six diffrent energies $E_K$ are shown 
Red circles show the amounts of neutron transfer, and the blue squares show those of proton transfer.
In a low energy case with $E_K =$ 2.50 MeV and $b$= 2, 3, 4, 5~fm, fusion appears.}
 \label{fig:4}
\end{figure}

The soliton existence is confirmed by calculating collision events systematically.
The fast charge equilibration mechanism, which is the generalized concept of fusion reaction, has been suggested to govern the mixing of protons and neutrons including fusion and deep inelastic collisions [Iw10p].
Under the appearance of fast charge equilibration, the mixing between protons and neutrons is known to take place quite rapidly within the order of 10$^{-22}$s [Iw10p] that should be compared to the typical duration time of low-energy nuclear reactions ($\sim$10$^{-20}$~s).
The charge equilibrating wave propagates at around 90$\%$ of the fermi velocity of many-nucleon systems (corresponding to the speed of zero sound propagation [Iw12]), so that the the propagation speed of charge equilibrating wave is roughly equal to a quarter of the speed of light. 
Soliton existence is false if we observe the charge equilibration. 
Consequently the soliton propagation is realized by the competition between the fast charge equilibration and the transparency originally due to a certain nonlinearity ($t_0$ and $t_3$ terms) of the TDDFT.
The fast charge equilibrating wave has been confirmed to play a role only if the collision energy is below the charge equilibration upper-limit energy.

In the lower energies less than a few MeV per nucleon, nuclear fusion appears and soliton cannot survive.
In the higher energies larger than 50 MeV per nucleon, nucleus breaks up into small pieces.
On the other hand, the fast charge equilibration wave can exist only below the upper-limit energy, where the upper energy is almost 80$\%$ of the fermi energy which is in accordance with the fact that the propagation speed is almost 90$\%$ of the fermi velocity. 
In case of nuclear collisions, this energy is roughly equal to 10 MeV per nucleon.
This fact may contain a hint to find out the soliton existence condition; i.e., it is reasonable to search for the energy just above the fast charge equilibration upper-limit energy.

\subsubsection{Numerical experiment}
Heuristic aspect of numerical experiment plays important roles in the past and present soliton theory (e.g., Fermi-Pasta-Ulam [Fe55]).
In this section systematic large-scale calculation of the nuclear collision dynamics is carried out based on the TDDFT.
Three-dimensional nuclear TDDFT calculations with the fully-introduced Skyrme-type interaction (10 parameters, as in the present calculations) are initiated in 1990's [Ki97].
On the other hand, many self-bound stationary states have already been calculated (theoretically found) by static calculations ($\partial \psi_i(t,{\bm r},\sigma) =0$ in the TDDFT) from 1980's, and they are compared to the experiments.
The impact parameter dependence of the soliton existence is systematically taken into account in three-dimensional calculations

Before moving on to the main discussion, we briefly review the preceding results [Iw15, Iw19]. 
According to the calculations dealing with 
$^{8}$He+$^{4}$He, 
$^{20}$O+$^{16}$O,
$^{44}$Ca+$^{40}$Ca,
$^{52}$Ca+$^{48}$Ca,
$^{104}$Sn+$^{100}$Sn,
$^{124}$Sn+$^{120}$Sn
reactions, the energy-dependence of soliton emergence has been clarified only for lighter cases: $^{8}$He+$^{4}$He, and $^{20}$O+$^{16}$O (cf. Fig. 4 of [Iw19]). 
For those lighter cases, a rough sketch of the energy-dependence is as follows: the soliton property is not so active for low energies less than a few MeV per nucleon, soliton property becomes active around 10 MeV per nucleon, it achieves almost the perfect transparency around 10–30 MeV per nucleon, and the transparency again decreases for much higher energies (Figs. 2 and 3 of [Iw19]).
For a mass dependence, the most decisive factor for the soliton propagation in heavier collisions has been clarified to be the appearance of the fragmentation including the nucleon emissions (mostly neutron emission).
On the other hand, massive momentum equilibration leading to the momentum equilibrium of each spatial point are activated around 80-100 MeV per nucleon, and those energies are too high to be relevant to the suppression of nuclear soliton propagation. 
In this article, by focusing on the stability of $N=Z$ nucleus of the two colliding nuclei, we clarify the energy-dependent soliton property of $^{4}$He and $^{16}$O.

The initial state of non-stationary problem is prepared by the two stationary solutions.
Let $A$ and $Z$ be mass number and the proton number of a colliding nucleus $^{A}Z$.
We consider a set of collisions:
\begin{equation} \begin{array}{ll}
\displaystyle ^{A}Z ~+~ ^{A+4}{Z} 
\label{eq27}
\end{array} \end{equation}
as a generalization of Eq.~(\ref{eq25}), where $(A,Z)=(4,2)$, $(16,8)$, $(40,20)$, $(48,20)$, $(100,50)$ and $(120,50)$ are taken into account.
Numerical solutions are obtained based on the finite difference method (for the details, see [Ma14]). 
Three-dimensional space is incremented by 1.0~fm, and the unit time step is set to one-third of 10$^{-23}$s. 
Vacuum boxes are prepared as $24 \times 24 \times 24$~fm$^3$ for the stationary problems, and as $64 \times 32 \times 32$~fm$^3$ for the nonstationary problems.
The center-of-mass of $^{A}Z$ and $^{A+4}Z$ are set to $(10,b/2,0)$ and $(-10,b/2,0)$, respectively, and the initial momentum of $^{A}Z$ and $^{A+4}{Z}$ to $(-\sqrt{2M_A E_K}, 0, 0)$ and $(\sqrt{2M_{A+4} E_K}, 0, 0)$, respectively.
The parameter $b$~fm imitates the impact parameter.
The quantities $M_A$ and $M_{A+4}$ denote the mass of  $^{A}Z$ and $^{A+4}{Z}$, respectively.
The periodic boundary condition is imposed in the three-dimensional Cartesian grid.

In Table \ref{table3} the binding energies of initial states are shown for confirming the quality of the present calculations.
The binding energy is not precisely the same as the experiment on the whole, but the difference is less than 15$\%$ for the lighter nuclei, and the difference less than 5$\%$ is achieved for heavier nuclei. 
It simply shows the quality of SV-bas parameter set.
By changing $A$, $Z$ and the two control parameters, we can examine the mass and energy dependence of the final products.
In particular, if the solitonic wave is dominant, no nucleon transfer takes place between $ ^{A}Z$ and $^{A+4}{Z}$.
If charge equilibrating wave is dominant, two neutron transfer from $ ^{A+4}Z$ to $^{A}{Z}$ is expected to be the most frequent reaction process.
For an astrophysical comparison it is practical to define the typical temperature of collision using the kinetic energy per nucleon or the relative velocity of the collision.
Based on the Bethe formula [Be37], the temperature of nuclear collision [Fr96, Iw19] is defined by
\begin{equation} \begin{array}{ll}
\displaystyle E_K = 
\left\{ \begin{array}{ll}
\kappa T_C T \qquad (T < T_C), \vspace{1.5mm} \\
\kappa T^2  \qquad  (T \ge T_C), 
\end{array} \right.
\label{eq28}
\end{array} \end{equation}
where $E_K$ is the total kinetic energy per nucleon, and $T_C = \kappa^{-1} = 7.2$~MeV is associated with the translation of the fermi energy of many-nucleon system to the relativistic center-of-mass kinetic energy [Iw13]. 
It shows that $E_K$ behaves linearly for low temperature and quadratically for high temperature.
In this article the results are shown by the kinetic energy $E_K = $1, 2, 3, $\cdots$, 10 MeV.

According to the previous study [Iw19], helium ($Z=2$) and oxygen ($Z=8$) isotopes have been proposed as the candidates of nuclear soliton.
This issue is examined from a stationary aspect.
For nuclei with $Z \le 20$, heavier nuclei become more stable than lighter nuclei, so that lighter nuclei tend to capture neutron or proton easily.
If this is also true to $^{4}$He and $^{16}$O, they cannot hold the soliton property.
For the verification of the proposed mechanism, the single neutron addition energy and single proton addition energy are approximately calculated using the energies of even-even nuclei.
From an energetic point of view, the following quantities are calculated.
\begin{equation} \begin{array}{ll} 
{\mathcal E}_n(A,Z) = \frac{ E(A+2,Z)  - E(A,Z) }{2} , \vspace{2.5mm} \\
{\mathcal E}_p(A,Z) =  \frac{E(A+2,Z+2)  - E(A, Z)}{2},  
\label{eq29}
\end{array} \end{equation}
where $E(A,Z)$ means the binding energy for the ground state of a nucleus consisting of $Z$ protons and $A-Z$ neutrons.
These quantities show the stability against adding one neutron (${\mathcal E}_n(A,Z)$) or one proton (${\mathcal E}_p(A,Z)$), respectively.
Neutron capture or proton capture is not preferred if the value is positive. 
The upper panel of Fig.~3 shows that the formation of density-functional field (a kind of mean-field) is not enough for $Z \le 8$ cases, and the directly-interacting few-body features are more important instead, where doubly-magic nuclei (helium and oxygen cases) show relatively good results comparable to experiments.
The stability of $^{4}$He and $^{16}$O can be found in the lower panel of Fig.~3.
A nucleus is stable against the addition of nucleons, if both ${\mathcal E}_n(A,Z)$ and ${\mathcal E}_p(A,Z)$ are positive.
$^{4}$He and $^{16}$O show the stability (lower panel of Fig.~3), although heavier cases with $Z \ge 10$ will find more stable bound system by adding neutrons.
From an experimental point of view,  $^{4}$He,  $^{12}$C and $^{16}$O are the candidate of soliton, where note that $^{8}$Be itself has known to be unbound system even before comparing to its neighbor nuclei.
From a theoretical and experimental point of view,  $^{4}$He and $^{16}$O are the candidates of soliton in which ${\mathcal E}_q(A,Z) $ values are positive. 
Consequently, the stability of soliton candidates $^{4}$He and $^{16}$O are confirmed with respect to the stability of stationary state in comparison to the neighbors.

\begin{table}
\caption{The observation probability $[\%]$ of soliton state of $^{4}$He calculated by ${\mathcal P}_m(E_K)$ and ${\mathcal P}_p(E_K)$ for given collision energies $E_K$.
Since the positions of center-of-mass are not so different for protons and neutrons in the present cases, the momentum transfer is shown as the total momentum transfer of all nucleons.
According to the preceding study, the soliton is suggested to exist around $E_K =$25.0~MeV [Iw19]. \\}
\begin{center} 
 ${\mathcal P}_m(E_K)$  \hspace{36mm}  ${\mathcal P}_p(E_K)$   \vspace{3mm}\\
\begin{tabular} {lrrrrrrrr}
$E_K$ [MeV]  & proton & neutron   \vspace{0.6mm} \\ \hline \vspace{-0.9mm} \\
2.50 \qquad  &  68.3 & 61.4  &  \\
5.00 \qquad &   75.2 & 69.4 &  \\
7.50 \qquad &   86.5 & 82.1 &  \\
10.0 \qquad &   90.7 & 95.0 &  \\
25.0 \qquad &   98.5 & 98.2 &   \\
  \hline 
\label{table3-1}
\end{tabular}
\qquad
\begin{tabular} {lrrrrrrrr}
$E_K$ [MeV]  & proton & neutron   \vspace{0.6mm} \\ \hline \vspace{-0.9mm} \\
2.50 \qquad  &  74.5 & 72.5  &  \\
5.00 \qquad &   60.2 & 56.2 &  \\
7.50 \qquad &   64.6 & 62.4 &  \\
10.0 \qquad &   76.9 & 73.7 &  \\
25.0 \qquad &   95.0 & 93.5 &   \\
  \hline 
\label{table3-2}
\end{tabular}
 \end{center}
\end{table}

A time evolution of $^{8}$He ~+~ $^{4}$He collision is shown in Fig.~4.
$^{8}$He is coming from the left hand side, and  $^{4}$He is moving from the right hand side.
It forms a rotating merged system around $t =28/3 \times 10^{-22}$s, and it is separated into two fragments as a result of collision.
As for Fig.~4,, the collision energy is selected as the lowest energy at which the solitonic wave component start to appear.
That is, taking at this energy as the standard energy,
for lower energies soliton cannot exist, and fusion reaction takes place;
for higher energies almost perfect transparency with respect to both mass and momentum is realized.
It is a non-central collision ($b\ne 0$) in which the axial symmetry along the collision axis is essentially violated.
The shape (more precisely, non-spherical property of the density and momentum distribution) is an important factor in multi-dimensional case, where less-symmetric shape of the merged system is introduced by the parameter $b$.
In addition to the internal excitation, a part of the total energy is delivered to the angular momentum of each nucleus in case of multi-dimensional and $b \ne 0$ cases.
To a certain degree, the appearance of rotational motion of the merged nucleus is a specific factor for the multi-dimensional soliton existence.

The detail of the nucleon transfer depends on the impact parameter, therefore on the shape and geometry.
For $^{8}$He~+~$^{4}$He collision transferred nucleon numbers are shown in Fig.~5, where the impact parameter dependence is shown in an energy-dependent manner.
By increasing the energy, nucleon transfer starts to disappear after $E_K =$7.50 MeV. 
Indeed, for cases with $E_K =$7.50, 10.0, 25.0, 50.0 MeV, the expectation value for the number of nucleon transfers are always less than 0.50, so that the soliton wave is concluded to be dominant in those cases.

With respect to the quantum mechanical observation, the calculated results are statistically summed up for a given collision energy (a given relative velocity).
Indeed, we cannot divide possible events by the impact parameter.
Using the concept of geometric cross section, the numbers of total cross section of all the inelastic events (events with touching between two nuclei) for a given collision energy $E_K$ is calculated by
\begin{equation} \begin{array}{ll} 
 \pi (1.50)^2  T(b_0,E_K)  +{\displaystyle \sum_{b_i =1}^{10} } (\pi (b_i +0.50)^2 - \pi (b_i -0.50)^2)  T(b_i,E_K) )  
\label{eq30}
\end{array} \end{equation}
where $b_i$~fm imitates the impact parameter; $T(b_i,E_K) = 1$ for touched cases, and $T(b_i,E_K) = 0$ for untouched case.
As readily understood by definition, events with large impact parameter hold larger cross section.
The rate of transparent events measured by the particle transparency for a given collision energy is calculated by
\begin{equation} \begin{array}{ll} 
{\mathcal P}_m (E_K) = 1 -
\frac{ \pi (1.50)^2 \left| N(b_0,E_K) ) \right|   + {\displaystyle \sum_{b_i=1}^{10} } (\pi (b_i+0.50)^2 - \pi (b_i-0.50)^2)  \left| N(b_i,E_K) ) \right| }
 {  \pi (1.50)^2  T(b_0,E_K)   + {\displaystyle \sum_{b_i=1}^{10} } (\pi (b_i+0.50)^2 - \pi (b_i-0.50)^2)  T(b_i,E_K) )    }  
\label{eq29}
\end{array} \end{equation}
where $N(b_i)$ is the transferred nucleon numbers.
This definition can be regarded as the probability, in which $|N(b_i,E_K)|$ is taken as 1 for $|N(b_i,E_K)| > 1$.
According to this treatment, ${\mathcal P}_m (E_K) $ can be regarded as the probability for the particle transparency.
Using the same definition using the geometric cross section, the transferred momentum rate is calculated. 
The rate of transparent events measured by the momentum transparency for a given collision energy is calculated by
\begin{equation} \begin{array}{ll} 
{\mathcal P}_p(E_K) = 1 -
\frac{ \pi (1.50)^2 \left| M(b_0,E_K) ) \right|   + {\displaystyle \sum_{b_i=1}^{10} } (\pi (b_i+0.50)^2 - \pi (b_i-0.50)^2)  \left| M(b_i,E_K) ) \right| }
 {  \pi (1.50)^2  T(b_0,E_K)   + {\displaystyle \sum_{b_i=1}^{10} } (\pi (b_i+0.50)^2 - \pi (b_i-0.50)^2)  T(b_i,E_K) )    }  
\label{eq29}
\end{array} \end{equation}
where $-1 \le M(b_i) \le 1$ is the transferred momentum divided by the initial momentum.

The soliton probability for all the possible collisions at given energies is summarized in Table~\ref{table3-2}.
For finding the soliton events at the energy just above the charge equilibration upper limit energy, it is reasonable to focus on  $E_K= 7.5$~MeV and  $E_K= 10.0$~MeV cases.
Indeed, for the reference case $E_K= 25$~MeV, single nucleon emission (neutron emission in most cases) is taking place during and after the collision, and the shapes are not well conserved.

In case of helium collisions, allmost 90$\%$ of the reaction is mass transparent for $E_K= 10.0$~MeV, and almost 80$\%$ is for $E_K= 7.5$~MeV.
The corresponding momentum transparency rate is 89.6 \% for $E_K =$10.0 MeV, and 92.1 \% for $E_K =$7.5 MeV. 
Consequently the probability for finding the soliton events being calculated by
\[ {\mathcal P}_p(E_K)  {\mathcal P}_p(E_K) \]
are 81 \% for $E_K =$10.0 MeV, and 76 \% for $E_K =$7.5 MeV.
In case of oxygen collisions, allmost 70 \% of the reaction is mass transparent for $E_K= 10.0$~MeV, and almost 60 \% is for $E_K= 7.5$~MeV.
The corresponding momentum transparency rate is 92.4 \% for $E_K =$10.0 MeV, and 87.8 \% for $E_K =$7.5 MeV. 
Consequently the probability for finding the soliton events are 65 \% for $E_K =$10.0 MeV, and 53 \% for $E_K =$7.5 MeV.
In both cases with helium and oxygen collisions, the cross section for soliton events is at the order of 1000 mb (milli-barn).
The soliton observation probabilities are larger than 50 \%, so that those collisions tend to be observed as the solitonic time-reversible events.

\section{Summary}   \label{sec5}
The soliton existence is nothing but the existence of perfect transparency, therefore the existence of perfect fluidity.
The theoretical evidence for the imperfect nuclear soliton existence has been presented for the first time in a realistic setting.
As a result ${^4}$He is concluded to be a candidate of nuclear soliton.
$^{16}$O also behaves like a soliton to a lesser degree.
As the fermi energy can be different for different fermions, the present study brings about a new insight on the validity of the different physics in different scales; through the competition relation, the existence of nuclear soliton has been shown to depend essentially on the fermi energy of many-nucleon systems.
An essential role of nonlinearity in the formation of  our material world is understood by the soliton propagation, since nucleon degree of freedom is related to the synthesis of chemical elements (H, He, Li, Be, $\cdots$). 
In conclusion, $^{4}$He and $^{16}$O are suggested to be the candidates for nuclear soliton.
From an applicational point of view, the soliton property of these nuclei will be utilized to the preservation of $^{4}$He matter, the condensation of $^{4}$He matter,  and production/synthesis of certain nucleus ( (by adding several $^{4}$He intentionally).

As seen in the competition mechanism between soliton wave propagation and charge equilibrating wave propagation,
the conditions for the soliton propagation depends essentially on the fermi energy of the fermionic quantum system.
Accordingly we have some conjectures to be confirmed in the near future.
In quantum systems,
\begin{enumerate}
\item the fermionic soliton exists in different scales in different ways, as the fermi energy is determined by fixing effective degrees of freedom;
\item the solitons between fermionic and bosonic systems are essentially different;
\item the general and special relativity effects changes the soliton existence;
\item another type of the soliton appears at the situation when fermions and bosons are tightly correlated (e.g. supersymmetric systems).
\item another type of the soliton exists in anyonic systems;
\end{enumerate}
where the first, second, and third conjectures are partly studied in this article.
The third and fifth conjectures are also associated with clarifying the difference compared to the maxwellian systems or anyonic systems.
That is, as in the present research, the soliton propagation in quantum systems should be examined by taking into account the spin degree of freedom and multi-dimensional spatial-degree of freedom.
The fourth conjecture is expected to play a role in clarifying and identifying the theory of everything.

As a closing remark some related open problems are pointed out.
Although there are several unknown and interesting topics in nuclear physics, we focus on the soliton propagation in many-nucleon systems.
\begin{itemize}
\item show the similarity/difference between the solitonic ``perfect fluidity'' and the bosonic ``superfluidity''.  
\item show quantitatively that the soliton propagation is suppressed/enhanced by the interaction terms other than $t_0$- and $t_3$-terms;
\item show the soliton existence probability under the influence of many-body dissipation (cf. one-body dissipation in the main text).
\item find the charge-parity symmetry breaking reaction in terms of the conditional/unconditional soliton existence (conditional/unconditional time-reversal symmetry).
\end{itemize}
These things will clarify the role of imperfect soliton in many-nucleon systems.
This kind of soliton should be different from the solitons in many-quark systems.


\section*{Acknowledgement}
This article is a transcript of the seminar ``The fundamentals of soliton theory and its application to many-nucleon systems'' at the department of physics, Kyoto University, Japan on Nov. 21, 2019.
Fruitful comments from Profs. K. Hagino, Y. Kanada-Enyo, N. Itagaki, K. Yoshida of Kyoto University and the other participants are acknowledged.
The author is grateful to Editors Profs. C. Simenel, P. Stevenson, D. Lacroix, L. Guo, and N. Schank for giving him a chance to write this article.
This work was partially supported by JSPS KAKENHI Grant No. 17K05440.
Numerical computation was carried out at Yukawa Institute Computer Facility of Kyoto University, and a workstation system at Kansai University.


\end{document}